\documentclass[]{article}
\usepackage{graphicx,amsmath,amssymb}
\def\d{{\rm d}}

\def\s{\sigma}
\def\del{\delta}
\def\al{\alpha}
\def\lam{\lambda}
\newcommand{\bl}[1]{\boldsymbol{#1}}

\title{\bf Critical index of Anderson transition \\in 3D systems}
\author{Arisato Kawabata\footnote{E-mail:
arisato.kawabata@gakushuin.ac.jp} \\
Department of Physics, Gakushuin University,\\ 1-5-1
Mejiro, Toshima-ku, Tokyo 171}
\begin{document}
\sloppy
\maketitle
\begin{abstract}
Anderson transition in three-dimensional systems is investigated
using renormalization group theory. $\beta$-function of a very simple
form is derived from a self-consistent consideration, and it gives a
value
$1+1/\sqrt{3}=1.58$ for the critical index of the transition, which is
very close to those obtained by numerical studies.
\end{abstract}

\noindent
keyword:

Anderson localization, metal-insulator transition, critical
index, renormalization group, self-consistent theory

\section{Introduction}

Metal-insulator transition of electronic systems due to randomness was
predicted by Anderson in 1958.~\cite{Ande} Let $x$ be a
quantity which controls the conductivity, such as impurity density or
electron density. Then the system undergoes a transition from a conductor
to an insulator at $x=x_c$, the critical value of $x$ (in the following
we assume that the system is metallic for
$x>x_c$). One of the main interests on this subject is the critical
behavior of the transition, namely, how the conductivity $\s(x)$
depends on $x$ near the critical point. In spite of an amount of
theoretical studies, little progress has been made about this
problem, until the appearance of the renormalization group (or
scaling) theory developed by Wegner~\cite{Wegn} and by
 Abrahams et al.~\cite{Abra} . 

The scaling hypothesis by Abrahams et al.
leads to the renormalization group equation~\cite{Kawa}  
\begin{equation}
\frac{\d \log g(L)}{\d\log L}=\beta(g(L))\,,
\label{rg}
\end{equation}
where $g(L)$ is the conductance of a system of size $L$.
The renormalization group theory predicts that the transition is
continuous in three-dimension:
$\s(x)$ vanishes continuously at $x=x_c$. The critical index (or exponent)
$\nu$ of the transition is defined by
the equation
\begin{equation}
\s(x)=C(x-x_c)^{\nu}\,.
\label{defnu}
\end{equation}
Here $C$ is a constant which is dependent on the microscopic
structure of the system, while $\nu$ is believed to be universal and
dependent only on the universality class, i.e., the fundamental
symmetry of the hamiltonian. Below we will treat the orthogonal
class, in which the Hamiltonian is composed of the kinetic energy
and the impurity potential, without spin-orbit coupling nor magnetic
field.

The critical index $\nu$ is given in terms of the $\beta$-function
$\beta(g)$:
\begin{equation}
\nu=\frac{1}{g_c \beta'(g_c)}\,,
\label{nu}
\end{equation}
where the prime on $\beta(g)$ indicates the derivative, and $g_c$ is the
fixed point value of $g$, satisfying
\begin{equation}
\beta(g_c)=0\,.
\label{gc}
\end{equation}

Elaborate calculations of the $\beta$-function have been done by
Wegner~\cite{Wegn2} and Hikami,~\cite{Hika} and their results lead to
$\nu\le 1$. This problem was investigated also by numerical methods.
Kramer and macKinnon developed a method to
estimate $\nu$ from numerical data, and they
obtained a value considerably larger than unity.~\cite{MacK,Kram}
Calculations with high accuracy have been performed recently by
Slevin and Ohtsuki, and they found that~\cite{Slev,Ohts}
\begin{equation}
\nu=1.57\sim1.58\,.
\label{numerical}
\end{equation}
 
Thus the discrepancy between the analytic and the numerical
methods is significant. At present, as long as the accuracy of the
results are concerned, the analytic methods seem less reliable than
the numerical methods. However, it is still important to pursue 
better analytic methods, because there are some aspects of which the
better understandings are obtained only by analytic methods.

In this paper we propose an analytic theory based on an idea
different from those of existing theories. We will find it well
reproduces the results of numerical studies.   
 
\section{Perturbational Calculation of $\beta$-function}

In this section we briefly review the method applied so far to calculate 
$\beta$-function, for it is the starting point of
the present theory, too. The details of the calculations are reviewed
in Ref.~\cite{Kawa} To be specific, we consider an electron gas system
with randomly distributed impurities. The hamiltonian is of the form
\begin{equation}
H = \frac{\bl{p}^2}{2m}+\sum_i u\,\del(\bl{r}-\bl{R}_i)\,,
\label{H}
\end{equation}
where $\bl{R}_i$'s are the positions of the impurities, which are assume
to be distributed uniformly with the average density $c_i$, independently
of each other.

So far the $\beta$-function was calculated in the form of the power series
in $1/g$. The term of $1/g$ arises from the lowest order interference
effect of electron wave scattered by impurities. Without
the interference correction, the conductivity is given by Boltzmann
(classical) conductivity
\begin{equation}
\s=\s_B\equiv\frac{n_e e^2\tau}{m}\,,
\label{sigmaB}
\end{equation}
where $n_e$ is the electron density, and $\tau$ is the electron scattering
time by impurities given by
\begin{equation}
\frac{1}{\tau}=\frac{2\pi}{\hbar}c_i u^2 N_0
\label{tau}
\end{equation}
$N_0$ being the density of states at the Fermi level. 

In thermal Green's function
formalism,~\cite{Abri} the interference correction to the Boltzmann
conductivity comes from the term corresponding to the Feymann
graph shown in Fig.~\ref{Fig1}. Here the solid lines and the dotted lines
indicate the one electron Green's functions and the impurity potential,
respectively; the wavy line indicates the cooperon propagator.
\begin{figure}[hbt]
\[\includegraphics[scale=0.65]{figure1.epsf}\]
\caption{The Feynmann graph for the lowest order interference correction}
\label{Fig1}
\end{figure}
Up to this order, the conductivity of an infinitely large system is given
by
\begin{equation}
\s=\s_B\left[1-\frac{1}{\pi\hbar N_0}\int\frac{\d
\bl{q}}{(2\pi)^3}\frac{1}{D_0 q^2}\right]\,,
\label{sigma}
\end{equation}
where $D_0$ is the diffusion coefficient defined by
\begin{equation}
D_0\equiv\frac{\s_B}{2e^2 N_0}=\frac{v_F^2\tau}{3}\,,
\label{D0}
\end{equation}
$v_F$ being the Fermi velocity, and the integral is to be done in the
region $q<q_c\approx 1/v_F\tau$. 

The conductance $g(L)$ is defined in terms of the conductivity $\s(L)$ of
the system of size $L$ as
\begin{equation}
g(L)=rL\s(L)\,,
\label{defgL}
\end{equation}
where $r$ is a constant of dimension of resistance to make $g(L)$
dimensionless, to be determined later. We postulate that $\s(L)$ is
given by the right hand side of  eq.~(\ref{sigma}) in which the
integral is cut off at
$q=\al/L$,
$\al$ being a numerical constant of order 1 (we consider the region
$\al/L<q_c$). Then we obtain
\begin{equation}
g(L)=rL\left[\s_B
-\frac{e^2}{\pi^3\hbar}\left(q_c-\frac{\al}{L}\right)
\right]\,,
\label{gL}
\end{equation}
and $\beta$-function is obtained from  eq.~(\ref{rg}):
\begin{equation}
\beta(g)=1-\frac{2}{g}\,,
\label{beta2}
\end{equation}
where we have taken $r$ so that
\begin{equation}
\frac{r\al e^2}{\pi^3\hbar}=2\,.
\label{cust}
\end{equation}
As is seen from  eqs.~(\ref{nu}) and (\ref{gc}), the ambiguity of
$r$ within a numerical factor does not affect the value of $\nu$, and we
find that
\begin{equation}
\nu=1
\label{nu2}\,.
\end{equation}

Calculations of $\beta$-function up to the order $1/g^4$ were done
by Wegner~\cite{Wegn2} and Hikami,~\cite{Hika}. Their $\beta$-functions
lead to $\nu=0.56$, and the deviation from the numerical value becomes
even larger. The inclusion of the higher order seems very difficult.
Moreover, it is not certain if the $1/g$ expansion is a proper
expansion. Thus we have to investigate the problem from a different
point of view.

\section{Self-Consistent Treatment}\label{self} 

Self-consistent treatments of Anderson transition were developed by
Vollhardt and Woelfle,~\cite{Voll} and by the present
author.~\cite{Kawa3} Although those theories contributed to the deeper
understanding of the problem, they give $\nu=1$ like the
renormalization group theory with the $\beta$-function (\ref{beta2}).
Those theories are based on the requirement of the consistency
between $\s$ and $D_0$ in the integrand (cooperon propagator) in 
eq.~(\ref{sigma}) for $L\to \infty$. In this paper we require the
consistency including the $L$ dependence of these quantities. 

In this sense  eq.~(\ref{sigma}) is not
self-consistent, for $\s(L)$ depends on $L$ while $D_0$ is a
constant. The diffusion coefficient must be independent of $q$
for small $q$, but it can depend on $q$ for large $q$.
Therefore, we assume that the diffusion coefficient behave like
$\propto q^\gamma$ for large $q$. We consider the case when $\al/L$ is
large in some sense. Then, assuming that the cooperon propagator
 is proportional to $1/q^{2+\gamma}$, and putting it into 
eq.~(\ref{sigma}) we find that
\begin{equation}
\s(L)\propto \left(\frac{\al}{L}\right)^{1-\gamma}\,,
\label{gamma}
\end{equation}
if we neglect the terms independent of $L$. We identify $\al/L$ with
$q$, and from the self-consistency requirement we find that
$\gamma=1/2$ and that the diffusion coefficient should behave like
$\sqrt{q}$ for large $q$. Scaling arguments predict a different $q$
dependence for large $q$,~\cite{chal} and this problem will be discussed
in the last section.

Thus, among various possibilities, we assume a simple form for the
cooperon propagator:
\begin{equation}
\frac{1}{D_0q^2(1+c\sqrt{\lam_0 q}) }\,,
\label{difp}
\end{equation}
where $c$ is a numerical constant to be determined self-consistently,
and
\begin{equation}
\lam_0\equiv \frac{1}{2\pi \hbar N_0 D_0}\,.
\label{lam0}
\end{equation}
The combination of $q$ with $\lam_0$ is based on the scaling
assumption that a length should be scaled by $\lam_0$. 

Then, by replacing the integrand in eq.~(\ref{sigma}) with 
the expression (\ref{difp}), we obtain 
\begin{align}
D(L)&\equiv \frac{\s(L)}{2e^2 N_0} \nonumber \\
&=D_0\Bigg[1+\frac{2}{\pi^2
c}\sqrt{\frac{\al \lam_0}{L}} -\frac{2}{\pi^2 c}\sqrt{\lam_0
q_c}+\frac{2}{\pi^2
c^2}\log\frac{c\sqrt{\lam_0q_c}+1}{c\sqrt{\al\lam_0/L}+1}\Bigg]\,.
\label{DL}
\end{align}

In comparing it with the denominator of the expression (\ref{difp}),
we neglect the last two terms in the square bracket. In fact, we will
see that the terms independent of $L$ does not affect the
renormalization group equation. Moreover, we have to compare them in
the regions $\sqrt{\al \lam_0/L} > 1$ and the $L$ dependence of the
logarithmic term is weaker than that of $\sqrt{\al \lam_0/L}$. Thus,
identifying $\al/L$ with $q$, the self-consistency requirement gives
\begin{equation}
c=\frac{2}{\pi^2 c}\,, \quad {\rm i.e.}, \quad c=\frac{\sqrt{2}}{\pi}\,.
\label{c}
\end{equation}
 
\section{Renormalization Group Equation}

We will derive a renormalization group equation from 
eq.~(\ref{DL}). Here we define
\begin{align}
g(L)&\equiv b\frac{L}{\lam(L)} \label{gL2}\,,\\
\lam(L) &\equiv \frac{1}{2\pi \hbar N_0 D(L)}\,,
\label{defgL2}
\end{align}
where $b$ is a numerical constant, and we easily find that this is
equivalent to  eq.~(\ref{defgL}). Then we obtain
\begin{align}
g(L)&=b\frac{L}{\lam_0}\left(1-\frac{\sqrt{2}}{\pi}\sqrt{\lam_0 q_c}
\right) \nonumber \\
&{}+b \frac{\sqrt{2}}{\pi}\sqrt{\al\frac{L}{\lam_0}}
+b\frac{L}{\lam_0}\log\frac{\sqrt{\lam_0
q_c}+\pi/\sqrt{2}}{\sqrt{\al\lam_0/L}+\pi/\sqrt{2}}\,,
\end{align}
and
\begin{equation}
\frac{\d\log g(L)}{\d \log
L}=\frac{1}{g(L)}\Bigg[g(L)-\eta\sqrt{b\frac{L}{\lam_0}}
+\frac{bL}{\lam_0}\frac{1}{2+\eta^{-1}\sqrt{bL/\lam_0}}\Bigg]\,,
\end{equation}
with $\eta\equiv \sqrt{b\al}/(\sqrt{2}\pi)$. 

We assume that the Anderson transition is described by
renormalization group equation. Then, the right hand side of this
equation have to be a universal function only of $g(L)$. From 
eqs.~(\ref{lam0}), (\ref{gL2}) and (\ref{defgL2}), we find that
$bL/\lam_0$ is $g(L)$ without the interference correction.
Therefore, it is reasonable to identify
$bL/\lam_0$ with exact $g(L)$. With this replacement, the
$\beta$-function is obtained:
\begin{equation}
\beta(g)=1-\frac{1}{\sqrt{g}}+\frac{1}{2+\sqrt{g}}\,,
\label{beta3}
\end{equation}
where we have chosen $b$ so that $\eta=1$.

\section{Critical Index}

The critical index $\nu$ is obtained from  eqs.~(\ref{nu}) and
~(\ref{gc}). Using  eq.~(\ref{beta3}) we easily find that
\begin{align}
g_c & =4-2\sqrt{3}\,,\\
\nu & =1+\frac{1}{\sqrt{3}}=1.58\,.
\end{align}
This value of $\nu$ agrees with that of numerical
studies,~\cite{Slev,Ohts} i.e., $\nu=1.57\sim1.58$.

\section{Summaries and Discussions}

On the basis of a self-consistent consideration, we have derived a
renormalization group equation for Anderson transition. It gives
 a critical index very close to those by numerical studies. We have not
intended to derive an exact result, and such a
good agreement is not the most important result of the present theory. 
It should be noted, however, that so far the analytic methods have never
given a critical index comparable with those by numerical
methods, within the author's knowledge, except for the rather
phenomenological theory by Shapiro.~\cite{Shap,Jans}

As regards the one parameter scaling, the recent numerical study by
Ohtsuki and Slevin revealed that the critical index is
little dependent on the definition of $g(L)$, i.e., what
kind of averaged conductance we take for it.~\cite{Slev2} This result is
very important because it indicates that a simple one parameter
renormalization group theory is reliable.

A crucial assumption in this theory is the form of the cooperon
propagator given by eq.~(\ref{difp}). As was mentioned in
Sec. \ref{self}, scaling arguments predict that the diffusion
coefficient $D(q)$ should depend on $q$ like $D(q)\propto q^{1+\eta'}$
with $\eta'\approx 1.3$ at the critical point, or for such $q$ that
$q\lam\gg 1$, where $\lam$ is the coherence length defined by 
eq.~(\ref{defgL2}).  On the other hand, from  eq.~(\ref{gL2}) we find
that $g\approx L/\lam \approx 1/(q\lam)$, and as regards the critical
behavior of the transition, the relevant values of $g$ are $g_c\approx
1$. Therefore, the relevant region of $q$ is $q\lam\approx 1$, and the
assumption $D(q) \propto \sqrt{q}$ in this region the does not
contradict to the scaling arguments.

Since the
$\beta$-function (\ref{beta3}) can not be obtained by the expansion in
$1/g$, probably it is impossible to derive the $\sqrt{q}$ term in the
expression (\ref{difp}) by perturbational approaches. The
$\beta$-function, however, reduces to eq.~(\ref{beta2}) for $g\gg 1$,
and it might be the reason why the
$1/g$ expansion seems to work for first order. For small $g$, on the
other hand, the $\beta$-function (\ref{beta3}) does not reduce to an
expected form $\beta(g)\approx \log g$, but, it does not invalidate the
present results, for, as was mentioned above, the important region of
$g$ is $g\approx 1$.

In this context, the
weak localization theory, which  was
very successful in explaining the magnetoresistance in fully metallic
regions~\cite{Kawa,HLN,Kawa4,Kawa5}, is not affected by the
introduction of $\sqrt{q}$ term into the cooperon propagator, for
the relevant scale of $q$ in that theory is
$1/\ell_B=\sqrt{eB/\hbar}$, $B$ being the magnetic flux density,
and is much smaller than $1/\lam_0$.

As for the experiments, many of them suggest that $\nu\approx
1$.~\cite{Kats}  However, recent careful analyses by Itoh
et al. indicate a possibility of $\nu>1$.~\cite{Itoh} From a
theoretical point of view, it is important to clarify the roles of
electron-electron interaction.

\section*{Acknowledgment}

This work is partly supported by "High Technology Research Center
Project" of Ministry of Education, Culture, Sports, Sciences and
Technology. The author is grateful to Prof. T. Ohtsuki for the
information on the present subject.

\end{document}